\documentclass{IEEEtran}
\usepackage[utf8]{inputenc}
\usepackage{fullpage}
\usepackage{amsmath,amssymb,url}
\usepackage{graphicx}

\usepackage{fullpage}
\usepackage{verbatim}
\usepackage{color}
\usepackage[operators]{cryptocode}
\usepackage{makecell}

\newtheorem{definition}{Definition}
\newtheorem{remark}{Remark}

\usepackage{tabularx}
\usepackage{booktabs}
\newcolumntype{Y}{>{\centering\arraybackslash}X}

\usepackage{bbm}

\def\Pr{\mathop{\rm Pr}\nolimits}

\DeclareMathOperator*{\argmax}{argmax}

\author{\IEEEauthorblockN{Kevin Atighehchi\IEEEauthorrefmark{1}, Loubna Ghammam\IEEEauthorrefmark{2}, Koray Karabina\IEEEauthorrefmark{3}, Patrick Lacharme\IEEEauthorrefmark{4}}

\IEEEauthorblockA{\IEEEauthorrefmark{1}Université Clermont Auvergne, LIMOS, France}

\IEEEauthorblockA{\IEEEauthorrefmark{2}ITK Engineering GmbH,  Germany}

\IEEEauthorblockA{\IEEEauthorrefmark{3}Florida Atlantic University, Boca Raton 33431, FL, United States}

\IEEEauthorblockA{\IEEEauthorrefmark{4}Normandie Univ, UNICAEN, ENSICAEN, CNRS, GREYC, 14000 Caen, France}
}

\title{A Cryptanalysis of Two Cancelable Biometric Schemes based on Index-of-Max Hashing}

\usepackage[normalem]{ulem}

\begin{document}

\maketitle

\begin{abstract}
Cancelable biometric schemes 
generate secure biometric templates by combining
user specific tokens and biometric data.
The main objective is to create irreversible, unlinkable, and revocable templates, with 
high accuracy of comparison.
In this paper, we cryptanalyze two recent 
cancelable biometric schemes based on a particular locality sensitive hashing function, index-of-max (IoM): Gaussian Random Projection-IoM (GRP-IoM) and Uniformly Random Permutation-IoM (URP-IoM). 
As originally proposed,
these schemes were claimed to be resistant against reversibility, authentication, and linkability attacks under the stolen token scenario.
We propose several attacks against GRP-IoM and URP-IoM, and argue that both schemes are severely vulnerable against authentication and linkability attacks. We also propose better, but not yet practical, reversibility attacks against GRP-IoM. 
The correctness and practical impact of our attacks are verified over the same dataset provided by the authors of these two schemes.
\end{abstract}


\section{Introduction}
\label{s: Introduction}

Biometrics has been widely adopted in
authentication systems, border control mechanisms, financial services,
and healthcare applications.
Biometric technologies are very promising 
to provide user-friendly, efficient, and secure 
solutions to practical problems.
In a typical biometric based authentication scheme, users register their biometric-related information with the system, and they are authenticated based on a similarity score calculated from their enrolled biometric data and the fresh biometric they provide. 
As a consequence, service providers need to manage biometric databases. This is somewhat analogous to storing and managing user passwords in a password-based authentication scheme. The main difference is that 
biometric data serves as a long-term and unique personal identifier, whence categorized as a highly sensitive and private data. This is not the case for passwords as they can be chosen independent of any user specific characteristics,
a single user can create an independent password per application, and passwords can be revoked, changed, and renewed easily at any time.
As a result, managing biometric data in applications is more challenging, and it requires more care. As biometric-based technologies are deployed at a larger scale,
biometric databases become natural targets in cyber attacks.
In order to mitigate security and privacy problems in the use of biometrics,
several biometric template protection methods have been proposed, including
cancelable biometrics, 
biometric cryptosystems (e.g.~fuzzy extractors),
keyed biometrics (e.g.~homomorphic encryption), 
and hybrid biometrics.
In this paper, we focus on cancelable biometrics (CB),
and refer the reader to two surveys \cite{Survey-2015,Survey-2016} for more details on biometric template protection methods.


In CB, a biometric template is computed through a process where the main inputs are biometric data (e.g.~biometric image, or the extracted feature vector) of a user, and a user specific token (e.g.~a random key, seed, or a password). 
In a nutshell, templates can be revoked, changed, and renewed by changing user specific tokens. For the security of the system, it is important that the template generation process is non-invertible (irreversible): given the biometric template and/or the token of a user, it should be computationally infeasible to recover any information about the underlying biometric data.
Similarly, given a pair of biometric templates and the corresponding tokens, it should be computationally infeasible to distinguish whether the templates were derived from the same user (unlinkability). 
We should note that even though user specific tokens in CB may be considered as secret, as part of a two-factor authentication scheme, cryptanalysis of CB with stronger adversarial models commonly assume that
the attacker knows both the biometric template and the token of a user. This is a plausible assumption in practice because a user token may have low entropy (e.g.~a weak password), or it may just be compromised by an attacker. This scenario is also known as the stolen-token scenario; see \cite{tkl08}.

CB was first propsed by Ratha \emph{et al.} \cite{rcb01} for face recognition. Since then, several CB schemes have been proposed, including the Biohashing algorithm applied on many modalities such as fingerprints \cite{tng04}, face \cite{tgn06}, and iris \cite{ctjnl06}.
CB schemes offer several advantages
such as efficient implementation, 
high accuracy of comparison,
and revocability. On the other hand, security of CB schemes, in general, are not well understood and the security claims are rather based on some intuitive, heuristic, and informal arguments, as opposed to being based on formal arguments with rigorous proofs.
As a matter of fact, several attacks on the CB schemes have been proposed; see
\cite{lcm09, LaChRo13, NaNaJa10, fly14, ToKaAzEr16} for attacks on biohashing type
schemes, and \cite{qfaf08,lh14}
for attacks on CB schemes using the Attack via Record Multiplicity (ARM) technique.

More recently, Jin et \textit{al.}~
\cite{Jin18Ranking} proposed two cancelable biometric schemes based on a particular locality sensitive hashing function, index-of-max (IoM) (see \cite{Charikar02Similarity} for details on IoM hashing): Gaussian Random Projection-IoM (GRP-IoM) and Uniformly Random Permutation-IoM (URP-IoM). 
It is shown in \cite{Jin18Ranking} that, for suitably chosen parameters, GRP-IoM and URP-IoM are robust against variation and noise in the measurement of data.
It is also claimed in \cite{Jin18Ranking} that, GRP-IoM and URP-IoM are resistant against reversibility, authentication, and linkability attacks under the stolen token scenario.

In this paper, we formalize some security notions under the stolen token scenario and propose several attacks against GRP-IoM and URP-IoM. We argue that both schemes are severely vulnerable against authentication and linkability attacks. We also propose better, but not yet practical, reversibility attacks for GRP-IoM. 
We utilize linear and geometric programming methods in our attacks. 
Linear programming has been previously used in the cryptanalysis of other schemes; see \cite{ToKaAzEr16}.
The correctness and practical impact of our attacks are verified over the same dataset provided by the authors of these two schemes. 
In order to be more specific, we state the security claims in \cite{Jin18Ranking} and our cryptanalysis results as follows:
\begin{enumerate}
    \item{\bf Reversibility attack:}
    In a reversibility attack, an adversary, who already has the knowledge of a user's specific token, and has at least one biometric template of the same user, tries to recover a feature vector, that corresponds to the user's biometric data.  
    \paragraph*{\bf Analysis in \cite{Jin18Ranking}}
    It is claimed in \cite{Jin18Ranking} that the best template reversing strategy for an adversary is to exhaustively search feature vectors. Based on some entropy analysis of the feature vectors, it is concluded in \cite{Jin18Ranking} that recovering the exact feature vectors from their 
    system implemented over the FVC 2002 DB1 dataset requires $2^{3588}$ operations for both GRP-IoM and URP-IoM; see Section~VII.A in \cite{Jin18Ranking}. In fact, the attack cost in \cite{Jin18Ranking} was underestimated as $(2^{12})^{299} = 2^{3588}$ because of underestimating $4636$ as $2^{12}$. A more accurate analysis yields a cost of $4636^{299} \approx 2^{3641}$.

    \paragraph*{\bf Our results} We propose a new reversibility attack against GRP-IoM. The main idea is to reduce the search space by guessing the sign of the components of the feature vectors with high success probability. Our analysis and experiments over the FVC 2002 DB1 dataset suggest that recovering GRP-IoM feature vectors now requires $2^{3466}$ operations. Even though our attack is not practical, it reduces the previously estimated security level for GRP-IoM by
    $3641-3433=208$ bits from $3641$-bit to $3433$-bit. Furthermore, we relax the {\it exact} reversibility notion to the {\it nearby} reversibility notion. This relaxation is reasonable given the fact that different measurements of the same user's biometric produce different feature vectors due to the inherent noise in the measurements. Under this relaxation, we propose successful attack strategies against GRP-IoM. Currently, we do not have any reversing attack strategy against URP-IoM that works better than the naive exhaustive search or random guessing strategies.
    For more details, please see Section~\ref{s: Rev GRP}. 
    
    \item{\bf Authentication attack:} In an authentication attack, an adversary, who already has the knowledge of a user's specific token, and has at least one biometric template of the same user, tries to generate a feature vector such that the adversary can now use that feature vector and the stolen token to be (falsely) authenticated by the system as a legitimate user. Note that authentication attacks are weaker than reversibility attacks because feature vectors generated in the attacks are not required to correspond to actual biometrics.
    \paragraph*{\bf Analysis in \cite{Jin18Ranking}} The authors in \cite{Jin18Ranking} analyze several authentication attack strategies (brute force, record multiplicity, false acceptance, birthday) against GRP-IoM and URP-IoM. In particular, the analysis in \cite{Jin18Ranking} yields that authentication attacks against GRP-IoM with parameters $m=300, q=16, \tau=0.06$, and URP-IoM with parameters $m=600, k=128, \tau=0.11$ require $2^{42}$ and $2^{252}$ operations, respectively, when the underlying dataset is FVC 2002 DB1; see Table~V in \cite{Jin18Ranking}.
    
    \paragraph*{\bf Our results} We utilize linear and geometric programming methods and propose new and practical authentication attacks against both GRP-IoM and URP-IoM. For example, we verify that our attacks against GRP-IoM and URP-IoM (under the same parameters and the dataset as above) run in the order of seconds and can authenticate adversaries successfully. We also show that the cancellability property of both GRP-IoM and URP-IoM are violated in the sense that adversaries can still be (falsely) authenticated by the system even after user templates are revoked and tokens are renewed. For more details, please see Section~\ref{s: Auth GRP} and~\ref{s: Auth URP}.

    \item{\bf Linkability attack:} In a linkability attack, an adversary, who is given a pair of biometric templates, tries to determine whether the templates were generated from two distinct individuals or from the same individual using two distinct tokens. 
    
    \paragraph*{\bf Analysis in \cite{Jin18Ranking}}
    Based on some experimental analysis of the pseudo-genuine and pseudo-imposter score distributions, and the large overlap between the two distributions, it is concluded in \cite{Jin18Ranking} that an adversary cannot be successful in a linkability attack against GRP-IoM and URP-IoM; see Section~VII~C in  \cite{Jin18Ranking}.
    
    \paragraph*{\bf Our results}
    Unlinkability claims in \cite{Jin18Ranking} are limited in the sense that the analysis only takes into account the attack strategies based on correlating the similarity scores of given templates. Therefore, the analysis in \cite{Jin18Ranking} does not rule out other, potentially better, attack strategies.
    In our analysis, we exploit partial reversibility of GRP-IoM and URP-IoM, and propose successful attack strategies (distinguishers) against both schemes.
    More specifically, the distinguisher for GRP-IoM uses a preimage finder along with a correlation metric, 
    that counts the number of identically signed components in the preimages. As a result, our attack can correctly link two templates 97 percent of the time. 
    The distinguisher for URP-IoM uses a preimage finder along with the Pearson correlation metric, and that can correctly link two templates 83 percent of the time. For more details, please see Section~\ref{s: linkability URP}.
\end{enumerate}

\paragraph*{\bf Organization}
The rest of this paper is organized as follows. We provide some background information on GRP-IoM and URP-IoM in Section~\ref{s: IoM algorithms}. In Section~\ref{s: IoM algorithms}, we also formalize some of the concepts for a more rigorous discussion and analysis of our attacks. We provide our attack models and relevant definitions in Section~\ref{s: attack_models}. Our attacks against GRP-IoM and URP-IoM are explicitly described and evaluated in Section~\ref{s: Attacks on GRP},~\ref{s: Attacking URP}, and~\ref{s: linkability URP}.
We derive our conclusions in
Section~\ref{s: Conclusion}.

\section{Formalizing Cancelable Biometric Schemes}
\label{s: IoM algorithms}

Biometric templates in GRP-IoM and URP-IoM are constructed in two steps: (1)~Feature extraction: A feature vector is derived from a biometric image; and (2)~Transformation: A user specific secret is used to transform the user's feature vector to a template.   
In this section, we present formal descriptions of these two steps and show how GRP-IoM and URP-IoM can be seen as concrete instantiations of our formal definitions. Our formalization will later help us to describe security notions, and to present our cryptanalysis of GRP-IoM and URP-IoM in a rigorous manner.

\subsection{Feature Extraction and Template Generation}

In the following, we let $(\mathcal{M}_A,D_A)$ and $(\mathcal{M}_B,D_B)$ be two metric spaces, where $\mathcal{M}_A$ and $\mathcal{M}_B$ represent the feature space and template space, respectively; and $D_A$ and $D_B$ are the respective distance functions.

\begin{definition}
A biometric feature extraction scheme is a pair of deterministic polynomial time algorithms $\Pi:=( E,V)$, where
\begin{itemize}
\item $E$ is the feature extractor of the system, that takes biometric data $b$ as input, and returns a feature vector $x \in \mathcal{M}_A$.
\item $V$ is the verifier of the system, that takes two feature vectors $x=E(b)$, ${x^\prime}=E(b')$, and a threshold $\tau_A$ as input, and returns $True$ if $D_A(x, {x^\prime}) \leq \tau_A$, and returns $False$ if $D_A(x, {x^\prime}) > \tau_A$.
\end{itemize}
\end{definition}
\begin{remark}
$V$ is not explicitly used in GRP-IoM and URP-IoM. More specifically, after a feature vector $x$ is extracted from a biometric image $b$, a transformation is applied to $x$ and a biometric template is derived. Therefore, the feature vector $x$ is not used in the protocol. The main reason that we introduce $V$ and $D_A$ here is to capture the notion of a vector $x^\prime$, that is {\it close} to the feature vector $x$. For example, the pair $x$ and $x^\prime$ may represent the feature vectors of the same individual extracted from two different measurements $b$ and $b^\prime$; in which case, one would expect $V$ to return $True$ for relatively small values of $\tau_A$. As a second example, $x^\prime$ may be the feature vector constructed by an attacker to reverse the biometric template of an individual with biometric image $b$. In this case, one may measure the success of the attack as a function of $\tau_A$, and the rate of $True$ values returned by $V$. A successful attack is expected to result in higher return rates of $True$ for relatively small values of $\tau_A$. 
\end{remark}

\begin{remark}
\label{s: Different similarity measures}
In this paper,
we consider two different methods
to quantify the similarity 
between feature vectors in the GRP-IoM and URP-IoM schemes.
The first one is 
the Euclidean distance,
where one computes 
\[d = D_{A}(x,{x^\prime})=\sqrt{\sum_{i=1}^{n}{(x_i-{x^\prime}_i)^2}},\]
and the verifier $V$ returns $True$ if $d \leq \tau_{Euc}$, and
returns $False$ if $d > \tau_{Euc}$.
We note that Euclidean distance is commonly deployed in biometric systems.
In the second method, one computes
\[s = S(x,{x^\prime})=\frac{\sum_{i=1}^n x_i \times {x^\prime}_i}{\sum_{i=1}^n x^2_i + {x^\prime}^2_i},\]
and the verifier $V$ returns $True$ if $s \ge \tau_{Sim}$, and
 returns $False$ if $s < \tau_{Sim}$.
The second method is a similarity score proposed in~\cite{Jin16Generating}, and deployed in the GRP-IoM and URP-IoM schemes in~\cite{Jin18Ranking}. 
One can see the close relation between the similarity score $s$ and the Euclidean distance $d$ through the equation
\[s = \frac{1}{2}\left(1 - \frac{d^2}{\sum_{i=1}^n x^2_i + {x^\prime}^2_i}\right).\]
In particular, the maximum value of $s$ is $1/2$, when 
$x=x^\prime$, or equivalently, when $d=0$.
One motivation for using this second method is the improved accuracy of the system. We refer the reader to \cite{Jin16Generating} for details.
\end{remark}




\begin{definition}
Let $\mathcal{K}$ be token (seed) space, representing the set of tokens to be assigned to users. A cancelable biometric scheme is a tuple of deterministic polynomial time algorithms $\Xi:=(\mathcal{G}, \mathcal{T}, \mathcal{V})$, where
\begin{itemize}
\item $\mathcal{G}$ is the secret parameter generator of the system, that takes a token (seed) $s\in \mathcal{K}$ as input, and returns a secret parameter set $sp$.
\item $\mathcal{T}$ is the transformation of the system, that takes a feature vector $x \in \mathcal{M}_A$, and the secret parameter set $sp$ as input, and returns a biometric template $u=\mathcal{T}(sp,x) \in \mathcal{M}_B$.
\item $\mathcal{V}$ is the verifier of the system, that takes two biometric templates $u$ = $\mathcal{T}(sp,x)$, ${u^\prime}=\mathcal{T}({sp^\prime},{x^\prime})$, and a threshold $\tau_B$ as input; and returns
$True$ if $D_B(u, {u^\prime}) \leq \tau_B$, and
returns $False$ if $D_B(u, {u^\prime}) > \tau_B$.
\end{itemize} 
\end{definition}

\subsection{GRP-IoM and URP-IoM Schemes}
\label{Algos_Spec}
The feature extractor $E$, which is common for both GRP-IoM and URP-IoM, takes fingerprint images as input, and generates feature vectors of length $299$, that is $\mathcal{M}_A=\mathbb{R}^{n}$ with $n=299$.

Let $I_a=\mathbb{Z}\cap [1,a]$ denote the set of integers from $1$ to $a$.
In \cite{Jin18Ranking}, GRP-IoM sets $\mathcal{M}_B = (I_q)^m$, and
URP-IoM sets
$\mathcal{M}_B = (I_k)^m$,
for some suitable parameters
$k$, $m$, and $q$. In the rest of this paper, we unify this notation and use $\mathcal{M}_B = (I_k)^m$ for both GRP-IoM and URP-IoM.
In both GRP-IoM and URP-IoM, the distance between two templates,
$D_B(u,u^\prime)$, is defined as the Hamming distance between $u$ and $u^\prime$. Therefore, in the rest of this paper, we use
$D_{\mathcal{H}}$ instead of $D_B$.

Both GRP-IoM and URP-IoM use an Index-of-Max operation, denoted $IoM$, in their verification algorithm $\mathcal{V}$. $IoM(v)$ is the smallest index, at which $v$ attains its maximum value. The algorithms 
$\mathcal{G}$ and $\mathcal{T}$ for GRP-IoM and URP-IoM significantly differ, and we explain them in the following.

\paragraph{\textbf{GRP-IoM Instantiation}}
\begin{itemize}
    \item $\mathcal{G}$ takes the seed $s$ as input, and generates random Gaussian \mbox{$n$-by-$k$} matrices $\mathcal{W}_i= [w_1 \cdots w_k]$, for $i=1, \ldots, m$. The column vectors of the matrices are sampled as 
    standard Gaussian vectors
    of length-$n$:
    $w_j \sample \mathcal{N}(0, I_n)$ for $j=1, \ldots, k$.
    As a result, the secret parameter set $sp$ consists of the sequence of projections $\mathcal{W}_1, \ldots, \mathcal{W}_m$.
    \item $\mathcal{T}$ takes the secret parameter set $\{\mathcal{W}_1, \ldots, \mathcal{W}_m\}$, and a fingerprint feature vector $x\in \mathbb{R}^n$ as input, and computes 
    \begin{enumerate}
     \item $v \gets x \mathcal{W}_i$,
     \item $u_i \gets IoM(v)$,
    \end{enumerate}
    for $i=1, \ldots, m$.
The output of $\mathcal{T}$ is the biometric template $u=(u_1, \ldots, u_m) \in (I_k)^m$.
 \item $\mathcal{V}$ takes two biometric templates $u,u^\prime\in(I_k)^m$, and 
 a threshold $0\le \tau\le 1$ as input; 
 computes $d = D_{\mathcal{H}}({u^\prime},u)$;
 and returns $True$ if $d/m \leq 1-\tau$, and
 returns $False$ if $d/m > 1-\tau$.
 Note that $\tau$ represents the minimum rate of the number of indices with the same entry in the pair of vectors to be accepted as a genuine pair.
\end{itemize}

\begin{figure}[h!]
\centering
\includegraphics[height=0.43\textwidth]{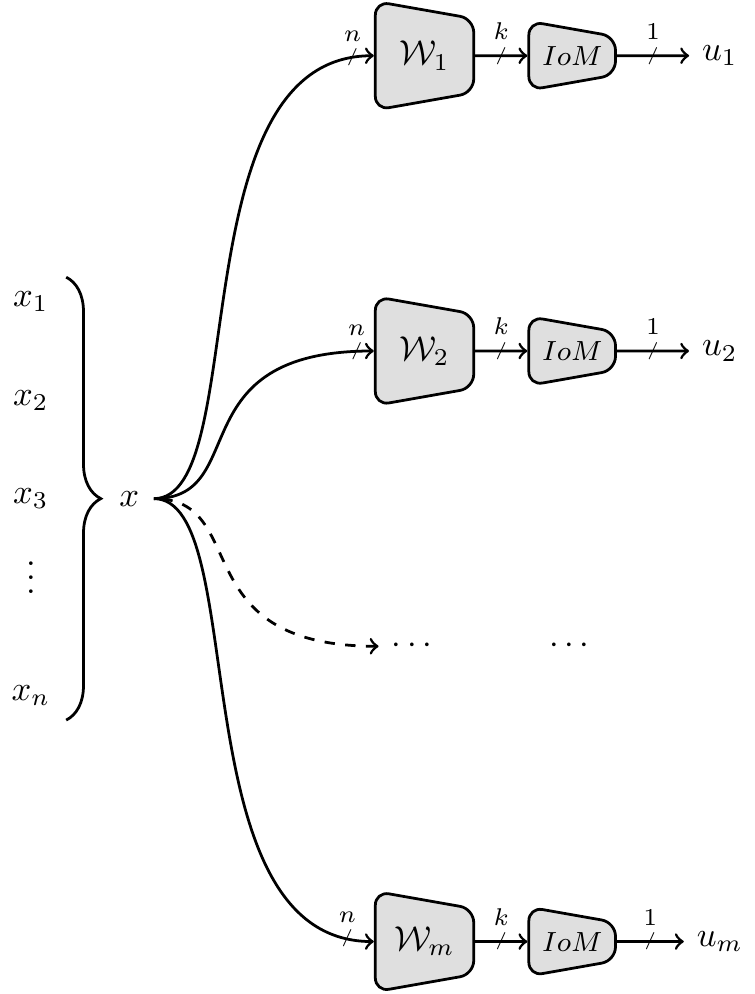}
\caption{Transformation of the GRP-based IoM scheme.}
\label{fig:IoM-GRP}
\end{figure}

The GRP-based IoM Hashing is depicted Figure~\ref{fig:IoM-GRP}.

\paragraph{Concrete parameters}
In \cite{Jin18Ranking}, several experiments are performed to select optimal parameters $k$ and $m$. More specifically, accuracy of the system is analyzed for $k\in\{2,3,5,10,50,100,150,200,250,299\}$, and $m\in \{2,5,10,50,100,150,200,250,300\}$. It is concluded that large $m$ is necessary for better accuracy, and that the effect of $k$ on the accuracy is not significant when $m$ is sufficiently large. For example, changing the configuration from $(k,m)=(2, 300)$ to $(k,m)=(250,300)$ changes the equal error rate (EER) of the system from $0.26\%$ to $0.24\%$, a minor improvement of $0.02\%$. As a result, the parameter set $(k,m)=(16,300)$ is commonly referred in the security and performance analysis of GRP-IoM in \cite{Jin18Ranking} with $\tau\in \{0.01,0.06\}$; see Table~IV and Table~V in \cite{Jin18Ranking}. 
For convenient comparison of our results, we also use $(k,m)=(16,300)$ and $\tau\in \{0.01,0.06\}$ as the main reference point in our security analysis in this paper.



\paragraph{\textbf{URP-IoM Instantiation}}

Let $S_n$ be the symmetric group of all permutations \mbox{$\sigma=(\sigma(1),\ldots, \sigma(n))$} on $(1, \ldots, n)$.
Let $S_{n,k}=\{\sigma=(\sigma(1),\ldots, \sigma(k)):\ \sigma\in S_n\}$ denote the set of partial permutations for 
$k \leq n$. In other words, permutations in $S_{n,k}$ are obtained by restricting permutations in $S_n$ to the first $k$ integers $1,2,...,k$.
For $\sigma\in S_{n,k}$ and $x=(x_1,...,x_n)$, we denote
$\sigma(x)=(x_{\sigma(1)},...,x_{\sigma(k)})$. 
As an example, for $n=5$ and $k=3$, restricting the permutation $(3,\ 5,\ 1,\ 2,\ 4)\in S_5$ to $S_{5,3}$ yields
$\sigma = (3,\ 5,\ 1)\in S_{5,3}$, and 
we get $\sigma((x_1,x_2,x_3,x_4,x_5)) = (x_3, x_5, x_1)$.
Finally, the component-wise (Hadamard) product of two vectors $x=(x_1,...,x_n)$ and $y=(y_1,...,y_n)$ is denoted by
$x\odot y = (x_1\cdot y_1,...,x_n\cdot y_n)$.
The secret parameter generation, transformation, and verification operations in URP-IoM are performed as follows:
\begin{itemize}
    \item $\mathcal{G}$ takes the seed $s$ as input, and generates partial permutations $\mathcal{P}_{ij}$ uniformly at random:
    $\mathcal{P}_{ij} \sample \mathcal{S}_{n,k}$ for $i=1, \ldots, m$, and $j=1,...,p$.
    As a result, the secret parameter set $sp$ consists of the sequence of partial permutations
    $\{(\mathcal{P}_{11}, \ldots,\mathcal{P}_{1p}), \ldots, (\mathcal{P}_{m1},\ldots,\mathcal{P}_{mp})\}$.
    
    \item $\mathcal{T}$ takes the secret parameter set $\{(\mathcal{P}_{11}, \ldots,\mathcal{P}_{1p}), \ldots, (\mathcal{P}_{m1},\ldots,\mathcal{P}_{mp})\}$, and a fingerprint feature vector $x\in \mathbb{R}^n$ as input, and computes
    \begin{enumerate}
        \item $v_j \gets \mathcal{P}_{ij}(x)$ for $j=1, \ldots, p$,
        \item $u_i \gets IoM(v_1 \odot \cdots \odot v_p)$,
    \end{enumerate}
for $i=1,...,m$.
The output of $\mathcal{T}$ is the biometric template $u=(u_1, \ldots, u_m) \in (I_k)^m$.

\item $\mathcal{V}$ takes two biometric templates $u,u^\prime\in(I_k)^m$, and 
 a threshold $0\le \tau\le 1$ as input; 
 computes $d = D_{\mathcal{H}}({u^\prime},u)$;
 and returns $True$ if $d/m \leq 1-\tau$, and
 returns $False$ if $d/m > 1-\tau$.
 Note that $\tau$ represents the minimum rate of the number of indices with the same entry in the pair of vectors to be accepted as a genuine pair.
\end{itemize}

\begin{figure}[h]
\centering
\includegraphics[height=0.45\textwidth]{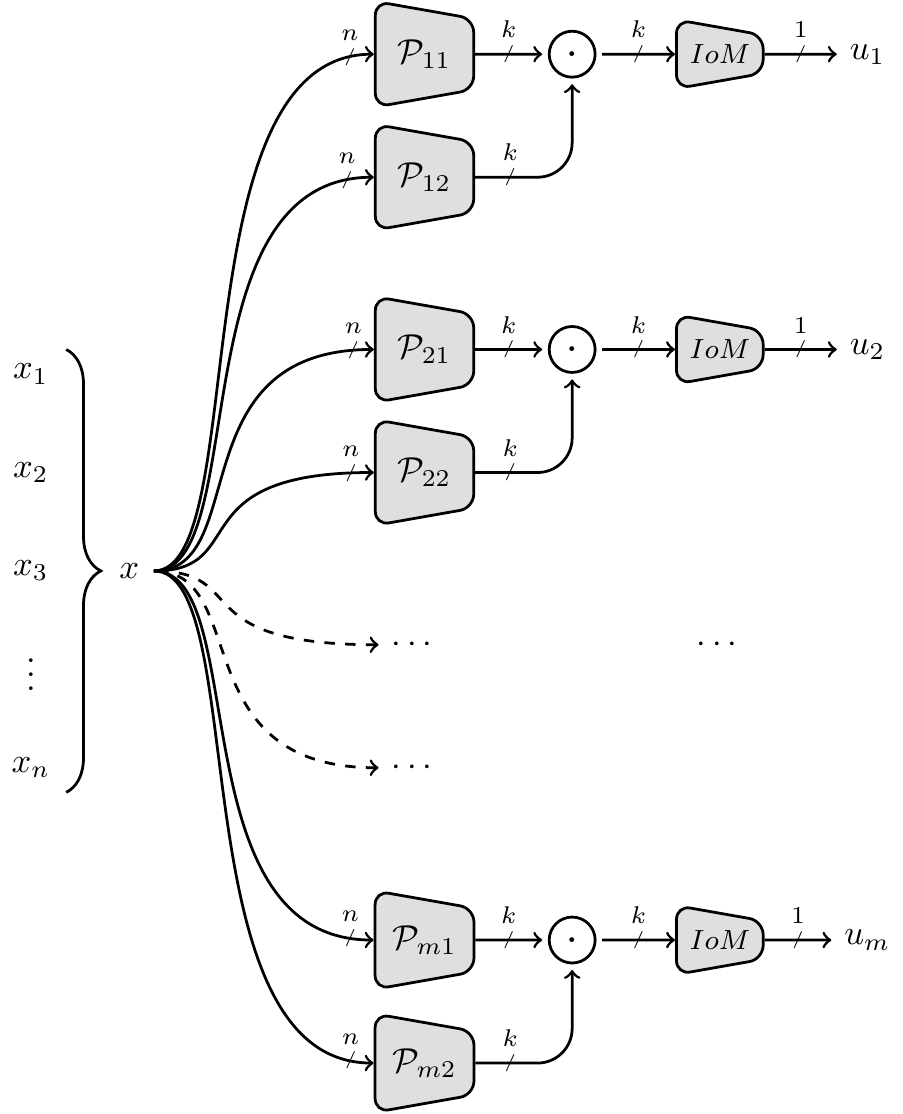}
\caption{Transformation of the URP-based IoM scheme for $p=2$.}
\label{fig:IoM-URP}
\end{figure}

An illustration of the URP-based IoM transformation is given Figure~\ref{fig:IoM-URP}.

\paragraph{Concrete parameters}
In \cite{Jin18Ranking}, several experiments are performed to select optimal parameters $k$, $m$, and $p$. It is reported that, the best performance over the FVC 2002 DB1 dataset is achieved when $(k,m,p)=(128, 600, 2)$ and $\tau=0.11$. This parameter set is also referred in the security and performance analysis of URP-IoM in \cite{Jin18Ranking}; see Table~V in \cite{Jin18Ranking}. For convenient comparison of our results, we also use $(k,m,p)=(128,600,2)$ and $\tau=0.11$ as the main reference point in our security analysis in this paper.

\label{fig: IoM-URP}



\section{Stolen Token Attack Models}
\label{s: attack_models}

Let $\mathcal{U}$ be the set of users of the biometric system. We identify a user with its biometric characteristic, and define a function $\mathcal{BC}(\cdot)$ that takes a biometric characteristic $usr \in \mathcal{U}$ as input, and outputs a digital representation of biometric data $b$; for instance, the scan of a fingerprint.
Note that for two different computations of 
$b=\mathcal{BC}(usr)$ and $b^\prime=\mathcal{BC}(usr)$ 
(e.g. at different times, or different devices), we may have
$b\ne b^\prime$ due to the inherent noise in the measurement
of biometric data. Therefore, we model 
$\mathcal{BC}(\cdot)$ as a probabilistic polynomial time function.
We also allow $\mathcal{BC}(usr_1) = \mathcal{BC}(usr_2)$
for $usr_1\ne usr_2$ due to the error rates of recognition systems.
In the following, we use $x \sample \mathcal{M}$ to indicate that $x$ is chosen from the set $\mathcal{M}$
uniformly at random.

\subsection{Reversibility attacks}

Let $x\in \mathcal{M}_A$ be a feature vector,
and let $u=\Xi.\mathcal{T}(sp,x)\in\mathcal{M}_B$ be the template generated from $x$ and the secret parameter set $sp$. In a reversibility attack, an adversary is given $u$, $sp$, and a threshold value $\tau_A$, and the adversary tries to find a feature vector $x^*\in \mathcal{M}_A$ such that $x^*$ is exactly the same as $x$, or $x^*$ is close to $x$ with respect to the distance function over $\mathcal{M}_A$ and the threshold value $\tau_A$.
In this case, we say that $x^*$ is a $\tau_A$-nearby-feature preimage (or simply a nearby-feature preimage, when $\tau_A$ is clear from the context) of the template $u$.
More formally, we have the following definition. 

\begin{definition} 
\textit{
Let $x\in \mathcal{M}_A$ be a feature vector,
and $u=\Xi.\mathcal{T}(sp,x)\in\mathcal{M}_B$ for some secret parameter set $sp$. Let $\tau_A$ be a threshold value.
A {\it nearby-feature preimage} of $u$ with respect to $sp$
is a feature vector $x^*$ such that
$\Pi.V(x,x^*,\tau_A)$ = $True$. 
}
\end{definition}

As a result, an adversary $\mathcal{A}$ in a reversibility attack can be modelled as an algorithm that 
takes $sp$ and $u$ = $\Xi.\mathcal{T}(sp,x)$ as input, and that outputs $x^* = \mathcal{A}(sp, u)\in\mathcal{M}_A$.  
We say that the adversary $\mathcal{A}$ is successful, if $x^*$ is a nearby-feature preimage of $u$.
More formally, we have the following definition.

\begin{definition}
\label{reversibility_success_rate_def}
\textit{
Let $\Xi$ be a cancelable biometric protection scheme and  $\mathcal{A}$ an adversary for a nearby-feature preimage attack. The success rate of $\mathcal{A}$, denoted by $\mathrm{Rate}^{Rev}_{\mathcal{A}}$, is defined as:
\[
\Pr\left[ \Pi.V(x,x^*,\tau_A) = True \;\middle|\;
  \begin{tabular}{@{}l@{}}
   $usr \sample \mathcal{U};\ s \sample \mathcal{K};$\\
   $b \gets \mathcal{BC}(usr);$\\
   $x \gets \Pi.E(b);$\\
   $sp \gets \Xi.\mathcal{G}(s)$;\\
   $u \gets \Xi.\mathcal{T}(sp,x);$\\ 
   $x^* \gets \mathcal{A}(sp, u);$
   \end{tabular}
  \right].
\]
}
\end{definition}

Note that an adversary can follow a {\it naive strategy} which consists in sampling a user $usr^*$ from $\mathcal{U}$ and returning $x^*=\mathcal{BC}(usr^*)$. Under this strategy, the adversary would be expected to succeed with probability $\mathrm{FMR}(\tau_A)$, which is the false accept rate of the system with respect to $\mathcal{U}$ and $\tau_A$ as the threshold value for the comparison of the pairs of feature vectors. A weakness of the scheme, 
with respect to the reversibility notion,
would require better attack strategies, and this motivates the following definition.

\begin{definition}
\label{reversibility_scheme_def}
\textit{
The protection scheme $\Xi$ is said to be reversible 
with advantage $\mathrm{Adv}^{Rev}(\mathcal{A})$, if there exists an adversary $\mathcal{A}$ such that $|\mathrm{Rate}^{Rev}_{\mathcal{A}} -  \mathrm{FMR}(\tau_A)|\ge \mathrm{Adv}^{Rev}(\mathcal{A})$.
If $\mathrm{Adv}^{Rev}(\mathcal{A})$ is negligible for all $\mathcal{A}$, then we say that $\Xi$ is irreversible in the stolen token scenario.}
\end{definition}

In particular, a protection scheme $\Xi$ is irreversible in the stolen token scenario if, the success rate of any adversary $\mathcal{A}$ is not significantly better than the success rate of the strategy of 
drawing $x^*$ randomly from $\mathcal{M}_A$.

\begin{remark}
Definitions~\ref{reversibility_success_rate_def} and~\ref{reversibility_scheme_def}
can be generalized to the case where the pair (token, template) is renewed $N$ times. The adversary thus takes advantage of $N$ pairs of (token, template), with $N>1$.
\end{remark}


\subsection{Authentication attacks}
Let $x\in \mathcal{M}_A$ be a feature vector,
and let $u=\Xi.\mathcal{T}(sp,x)\in\mathcal{M}_B$ be the template generated from $x$ and the secret parameter set $sp$. In an authentication attack, an adversary is given $u$, $sp$, and a threshold value $\tau_B$, and the adversary tries to find a feature vector $x^*\in \mathcal{M}_A$ such that 
for $u^*=\Xi.\mathcal{T}(sp,x^*)$, 
$u^*$ is exactly the same as $u$,
or $u^*$ is close to $u$ with respect
to the distance function over $\mathcal{M}_B$ and the threshold value $\tau_B$.
In this case, we say that $x^*$ is a $\tau_B$-nearby-template preimage (or simply a nearby-template preimage, when $\tau_B$ is clear from the context) of the template $u$.
More formally, we have the following definition. 
\begin{definition}
\textit{
Let $x\in \mathcal{M}_A$ be a feature vector,
and $u=\Xi.\mathcal{T}(sp,x)\in\mathcal{M}_B$ for some secret parameter set $sp$. Let $\tau_B$ be a threshold value.
A {\it nearby-template preimage} of $u$ with respect to $sp$
is a feature vector $x^*$ such that
$u^*=\Xi.\mathcal{T}(sp,x^*)$ and
$\Xi.\mathcal{V}(u,u^*,\tau_B) = True$.
}
\end{definition}

As a result, an adversary $\mathcal{A}$ in an authentication attack can be modelled as an algorithm that 
takes $sp$ and $u$ = $\Xi.\mathcal{T}(sp,x)$ as input, and that outputs $x^* = \mathcal{A}(sp, u)\in\mathcal{M}_A$.  
We say that the adversary $\mathcal{A}$ is successful if $x^*$ is a nearby-template preimage of $u$.
More formally, we have the following definition.

\begin{definition}
\textit{
Let $\Xi$ be a cancelable biometric protection scheme and  $\mathcal{A}$ an adversary for finding a nearby-template preimage.
 The success rate of $\mathcal{A}$, denoted by $\mathrm{Rate}^{Auth}_{\mathcal{A}}$, is defined as:
\[
\Pr\left[ \Xi.\mathcal{V}(u,u^*,\tau_B) = True \;\middle|\;
  \begin{tabular}{@{}l@{}}
   $usr \sample \mathcal{U};\ s \sample \mathcal{K};$\\
   $b \gets \mathcal{BC}(usr);$\\
   $x \gets \Pi.E(b);$\\
   $sp \gets \Xi.\mathcal{G}(s)$;\\
   $u \gets \Xi.\mathcal{T}(sp,x);$\\ 
   $x^* \gets \mathcal{A}(sp, u);$\\
   $u^* \gets \Xi.\mathcal{T}(sp,x^*);$
   \end{tabular}
  \right].
\]
}
\end{definition}

Note that an adversary can follow 
the aforementioned {\it naive strategy}.
Under this strategy, the adversary would be expected to succeed with probability $\mathrm{FMR}(\tau_B)$, which is the false accept rate of the system with respect to $\mathcal{U}$ and $\tau_B$ as the threshold value for the comparison of the pairs of templates. This strategy is also commonly known as the false match (or acceptance) rate attack in the literature. A weakness of the scheme, with respect to the false authentication notion, would require better attack strategies, and this motivates the following definition.

\begin{definition}
\textit{
The protection scheme $\Xi$ is said to have false authentication with advantage $\mathrm{Adv}^{Auth}(\mathcal{A})$ property, if there exists an adversary $\mathcal{A}$ such that $|\mathrm{Rate}^{Auth}_{\mathcal{A}} -  \mathrm{FMR}(\tau_B)|\ge \mathrm{Adv}^{Auth}(\mathcal{A})$.
If $\mathrm{Adv}^{Auth}(\mathcal{A})$ is negligible for all $\mathcal{A}$, then we say that $\Xi$ does not have false authentication property under the stolen token scenario.}
\end{definition}

In particular, a protection scheme $\Xi$ does not have false authentication property under the stolen token scenario, if the success rate of any adversary $\mathcal{A}$ is not significantly better than the success rate of the strategy of 
drawing $x^*$ randomly from $\mathcal{M}_A$; or in other words, the success rate of any attack is bounded by the false match rate of the system.

Now, suppose that an adversary
knows the secret parameter set 
$sp$ of a user ($usr$), and the template
$u=\Xi.\mathcal{T}(sp,x)$ of the user, where
$x=\Pi.E(\mathcal{BC}(usr))$.
At this point,  the user may renew her token, or register to another system with a new token and a freshly acquired feature vector.
Suppose now that the adversary knows the 
user's new secret parameter set ${sp^\prime}$, but
the adversary does not know the user's new template ${u^\prime}=\Xi.\mathcal{T}({sp^\prime},{x^\prime})$. In such a scenario, the adversary would try to compute
a nearby-template preimage $x^*$ of 
the template ${u^\prime}=\Xi.\mathcal{T}({sp^\prime},{x^\prime})$,
given $sp$, $u$, and ${sp^\prime}$.
Informally, we call such a nearby-template preimage $x^*$ as a \emph{long-lived nearby-template preimage}. More formally, we have the following
definition.
\begin{definition}
\label{ll_preimage_finder}
\textit{
Let $\Xi$ be a cancelable biometric protection scheme and  $\mathcal{A}$ an adversary for finding a long-lived nearby-template preimage.
 The success rate of $\mathcal{A}$, denoted by $\mathrm{Rate}^{Auth\mbox{-}\ell\ell}_{\mathcal{A}}$, is defined as:
\[
\Pr\left[ 
  \begin{tabular}{@{}c@{}}
  $\Xi.\mathcal{V}({u^\prime},{u^\prime}^*,\tau_B) = True$
  \end{tabular}
 \;\middle|\;
  \begin{tabular}{@{}l@{}}
   $usr \sample \mathcal{U};$\\ 
   $b \gets \mathcal{BC}(usr);$\\ 
   $b' \gets \mathcal{BC}(usr);$\\ 
   $x \gets \Pi.E(b);$\\
   ${x^\prime} \gets \Pi.E(b');$\\
   $s \sample \mathcal{K}; s' \sample \mathcal{K};$\\
   $sp \gets \Xi.\mathcal{G}(s);$\\
   ${sp^\prime} \gets \Xi.\mathcal{G}(s')$;\\
   $u \gets \Xi.\mathcal{T}(sp,x);$\\ 
   ${u^\prime} \gets \Xi.\mathcal{T}({sp^\prime},{x^\prime});$\\ 
   $x^* \gets \mathcal{A}(sp, {sp^\prime}, u);$\\
   ${{u^\prime}}^* \gets \Xi.\mathcal{T}({sp^\prime},x^*);$
   \end{tabular}
  \right].
\]
}
\end{definition}

Again, an adversary can follow 
the aforementioned {\it naive strategy}.
Under this strategy, the adversary would be expected to succeed with probability $\mathrm{FMR}(\tau_B)$, as explained in the previous authentication attack model.
A weakness of the scheme, with respect to the long-lived false authentication notion, would require better attack strategies, and this motivates the following definition.

\begin{definition}
\label{ll_preimage_finder-adv}
\textit{
The protection scheme $\Xi$ is said to have long-lived false authentication
with advantage $\mathrm{Adv}^{Auth-\ell\ell}(\mathcal{A})$ property, if there exists an adversary $\mathcal{A}$ such that $|\mathrm{Rate}^{Auth-\ell\ell}_{\mathcal{A}} -  \mathrm{FMR}(\tau_B)|\ge \mathrm{Adv}^{Auth-\ell\ell}(\mathcal{A})$.
If $\mathrm{Adv}^{Auth-\ell\ell}(\mathcal{A})$ is negligible for all $\mathcal{A}$, then we say that $\Xi$ does not have long-lived false authentication property under the stolen token scenario.}
\end{definition}

In other words, a protection scheme $\Xi$ is vulnerable to long-lived nearby-template preimage attacks if an adversary, who knows a user's previous token and template pair, and the user's renewed token, can construct a feature vector that can be (falsely) authenticated by the system with some probability greater than the false accept rate of the system.

\begin{remark}
We should emphasize that
in finding long-lived nearby-template preimages,
we allow the adversary to know $sp$, $u$, and ${sp^\prime}$, but
we do not allow the adversary to know ${u^\prime}$. Therefore, the finding a long-lived nearby-template preimage
problem is not easier than 
the finding a nearby-template preimage problem. This observation also makes sense in practice as explained in the following. Consider an adversary, who has access to 
an efficient algorithm for finding nearby-template preimages. Such an adversary can be blocked 
by revoking biometric templates and renewing tokens.
On the other hand, an adversary, who has access to an efficient algorithm for finding long-lived nearby-template preimages,
can still be (falsely) authenticated by the system
even after user templates are revoked and tokens are renewed. In other words, a successful algorithm for finding long-lived nearby-template preimages would defeat the purpose of
cancellability feature of a system.
\end{remark}

\begin{remark}
Definitions~\ref{ll_preimage_finder} and~\ref{ll_preimage_finder-adv}
can be generalized to the case where the pair (token, template) is renewed $N$ times. The adversary thus takes advantage of the $N$ first leaked pairs of (token, template), along with the $(N+1)$'th token.
\end{remark}

\subsection{Linkability attacks}
Let $x,{x^\prime}\in\mathcal{M}_A$ be two feature vectors. 
Let $u=\Xi.\mathcal{T}(sp,x)\in\mathcal{M}_B$ and 
${u^\prime}=\Xi.\mathcal{T}({sp^\prime},{x^\prime})\in\mathcal{M}_B$
be two templates generated from $x$ and ${x^\prime}$, and the secret parameters set $sp$
and ${sp^\prime}$.
In a linkability attack, an adversary is given
$sp$, ${sp^\prime}$, $u$, and ${u^\prime}$,
and the adversary tries to find out whether $x$ and ${x^\prime}$ are derived 
from the same user.
As a result, an adversary $\mathcal{A}$ in a linkability attack can be modelled as an algorithm that 
takes $sp$, ${sp^\prime}$, $u$, and ${u^\prime}$ as input, and that outputs $0$ or $1$,
where the output $1$ indicates that 
the feature vectors $x$ and ${x^\prime}$ are extracted from the same user, and
the output $0$ indicates that 
the feature vectors $x$ and ${x^\prime}$ are extracted from two different users.
We say that the adversary $\mathcal{A}$ is successful, if his conclusion (whether the feature vectors are extracted from the same user)
is indeed correct.
More formally, we have the following definition.

\begin{definition}
\textit{
Let $\Xi$ be a cancelable biometric protection scheme and  $\mathcal{A}$ an adversary for a linkability attack.
The success rate of $\mathcal{A}$, denoted by $\mathrm{Rate}^{Link}_{\mathcal{A}}$, is defined as:
\[
\Pr\left[ c' = c \;\middle|\;
  \begin{tabular}{@{}l@{}}
   $usr \sample \mathcal{U};$\\ 
   $b \gets \mathcal{BC}(usr);\ x \gets \Pi.E(b);$\\
   $s \sample \mathcal{K}; sp \gets \Xi.\mathcal{G}(s)$;\\
   $s' \sample \mathcal{K}; {sp^\prime} \gets \Xi.\mathcal{G}(s')$;\\
   $c \sample \{0,1\}$;\\
   $usr' \leftarrow usr$  if $c=0$;\\
   $usr' \sample \mathcal{U}\setminus usr$ if $c=1$;\\
   $b' \sample \mathcal{BC}(usr');\ {x^\prime} \gets \Pi.E(b');$\\
   $u \gets \Xi.\mathcal{T}(sp,x); {u^\prime} \gets \Xi.\mathcal{T}({sp^\prime},{x^\prime});$\\ 
   $c' \gets \mathcal{A}(sp, u, {sp^\prime}, {u^\prime});$
   \end{tabular}
  \right].
\]
}
\end{definition}

Note that an adversary can follow a naive strategy by simply 
sampling a value from $\{0,1\}$ uniformly at random.
Under this strategy, the adversary would be expected to succeed with probability $1/2$. This strategy is also known as the guessing attack in the literature.
A weakness of the scheme, with respect to the linkability notion, would require better attack strategies, and this motivates the following definition.

\begin{definition}
\label{linkability-adv}
\textit{
The protection scheme $\Xi$ is said to be linkable (distinguishable)
with advantage $\mathrm{Adv}^{Link}(\mathcal{A})$, if there exists an adversary $\mathcal{A}$ such that $|\mathrm{Rate}^{Link}_{\mathcal{A}} -  1/2|\ge \mathrm{Adv}^{Link}(\mathcal{A})$.
If $\mathrm{Adv}^{Link}(\mathcal{A})$ is negligible for all $\mathcal{A}$, then we say that $\Xi$ is unlinkable (indistinguishable) under the stolen token scenario.}
\end{definition}

We should note that our linkability definitions are aligned with  the definitions in \cite{Unlinkability}.
In particular, 
the \textit{gradual linkability} (e.g.~fully linkable, linkable to a certain degree) concepts as described in \cite{Unlinkability} are captured through
measuring the success rate of an adversary in our framework.

\section{Attacks on GRP-IoM}
\label{s: Attacks on GRP}

In this section, we propose some concrete attack strategies against
GRP-IoM, and evaluate the impact of our attacks through our implementation over one of the datasets as provided in \cite{Jin18Ranking}.
More specifically, we use the dataset of features 
extracted 
from the fingerprint images of FVC2002-DB1
as in~\cite{Jin18Ranking}. 
This dataset contains a total of 500 samples:
5 fixed-length feature vectors per user; for a total of 100 users.
A feature vector is derived from a fingerprint image in two steps: minutiae descriptor extraction (MCC), followed by a kernel learning-based  transformation; see~\cite{Jin18Ranking,Jin16Generating} for more details.

As mentioned before,
for convenient comparison of our results, we use the GRP-IoM paramaters $n=299$, $(k,m)=(16,300)$ and $\tau\in \{0.01,0.06\}$ as the main reference point in our security analysis, because these parameters are commonly referred in the security and performance analysis of GRP-IoM; see Table~IV and Table~V in \cite{Jin18Ranking}.

\subsection{Authentication attacks on GRP-IoM}
\label{s: Auth GRP}

\paragraph*{\bf Finding nearby-template preimages} As before, let $x\in \mathcal{M}_A=\mathbb{R}^{n}$ be a feature vector,
and let $u=\Xi.\mathcal{T}(sp,x)\in\mathcal{M}_B=\mathbb{R}^{m}$ be the template generated from $x$ and the secret parameter set $sp$. Assume that an adversary $\mathcal{A}$ knows $u$ and $sp$.
In order to find a nearby-template preimage vector $x^*\in \mathbb{R}^{n}$, the adversary proceeds as follows.
Since $\mathcal{A}$ knows $sp$, $\mathcal{A}$ can recover 
the set of Gaussian random projections in GRP-IoM:
$A=\{\mathcal{W}_i \in \mathbb{R}^{k\times n}, 1 \leq i \leq m\}$  Let the rows of $\mathcal{W}_i$ be denoted by $\mathcal{W}_{i,1}$, $\mathcal{W}_{i,2}$, ..., $\mathcal{W}_{i,k}$.
Let $\langle \mathcal{W}_{i,j}, x \rangle$ denote the inner product between the vectors $\mathcal{W}_{i,j}$ and $x$.
Recall that the template produced by GRP-IoM is a vector $u=(u_1, u_2, \ldots, u_m) \in (\mathbb{Z}_k)^m$ comprised of the indices of maximum, \emph{i.e.} 
\[u_i=\argmax_{j\ \textrm{s.t.}\ 1 \leq j \leq k}  \langle \mathcal{W}_{i,j},x \rangle\ \textrm{for}\ i=1 \ldots m,\]
from which $\mathcal{A}$ recovers the set of inequalities 
\begin{equation} 
\label{GRP_full_set_constraints}
\{ \langle \mathcal{W}_{i,j}, {x^\prime} \rangle \leq \langle \mathcal{W}_{i,u_i}, {x^\prime} \rangle \}_{1 \leq i \leq m, 1 \leq j \leq k}.
\end{equation}

As a result,  $\mathcal{A}$ obtains $(k-1)m$ inequalities in $n$ unknowns, and 
sets $x^*$ to be one of the (arbitrary)
solutions of this system
(possibly imposing $|x^*_i| \leq c$ for some $c$ positive, for $1 \leq i \leq m$).
By the construction of 
$x^*$, we must have
$u^* = \Xi.\mathcal{T}(sp,x^*)=\Xi.\mathcal{T}(sp,x) = u$, and so 
$\Xi.\mathcal{V}(u,u^*,\tau_B) = True$, for all $\tau_B$.
In other words, $\mathcal{A}$ is expected to get (falsely) authenticated by the server with $100\%$, or equivalently, $\mathrm{Rate}^{Auth}_{\mathcal{A}}=100\%$.
The expected success rate of our attack
has been verified in our python implementation 
using the CVXOPT library~\cite{cvxopt}
on a computer 
running on Ubuntu 17.10 with Xfce environment,
with a Core i7 4790k 4Ghz processor, 8GB of RAM, and a SATA SSD of 512GB. The attack runs in the order of seconds for the parameters $n=299$, $(k,m)=(16,300)$ and $\tau\in \{0.01,0.06\}$. 

\paragraph*{\bf Finding long-lived nearby-template preimages}
Let $x$ and ${x^\prime}$ be two feature vectors
of the same user, and $sp$ and ${sp^\prime}$
two secret parameters sets.
Let $u=\Xi.\mathcal{T}(sp,x)$ and
${u^\prime}=\Xi.\mathcal{T}({sp^\prime},{x^\prime})$.
In finding a long-lived nearby-template preimage $x^*$ of ${u^\prime}$, we assume that the adversary
$\mathcal{A}$ knows $u$, $sp$, ${sp^\prime}$.
In our proposed attack, $\mathcal{A}$ follows the previously described strategy to find a nearby-template preimage $x^*$ based on $u$ and $sp$, and presents this
$x^*$ as a candidate for nearby-template preimage of ${u^\prime}$.

We evaluate this attack by computing both the average and the minimum comparison score, over one hundred users, between ${{u^\prime}}^*=\Xi.\mathcal{T}({sp^\prime},x^*)$ and the re-enrolled genuine template ${u^\prime}=\Xi.\mathcal{T}({sp^\prime},{x^\prime})$. 
Our experiments yield $44.6\%$ as the  average rate of the number of indices with the same entry in ${u^\prime}$ and ${u^\prime}^*$;
and $17.7\%$ as the minimum rate of the number of indices with the same entry in ${u^\prime}$ and ${u^\prime}^*$. Therefore, given the comparison score thresholds of $\tau\in\{0.01,0.06\}$ as set in~\cite{Jin18Ranking},
we expect that
the success rate of the adversary to be
$\mathrm{Rate}^{Auth-\ell\ell}_{\mathcal{A}}=100\%$.
The above attack strategies show that GRP-IoM is severely vulnerable against authentication attacks under the stolen token and template attack model, and also show that adversaries cannot be prevented by renewing templates or tokens. In other words, the cancellability feature of GRP-IoM is violated under the stolen token and template scenario.

\paragraph*{\bf Optimizing authentication attacks}
Next, we explore whether the attacks can be optimized when a user leaks several token and template pairs.
More specifically, assume that an adversary
captures $N$ token, template pairs 
$(sp_i,u^{(i)})$, for $1 \leq i \leq N$, derived from distinct feature vectors of the same user. In practice, templates are derived from feature vectors extracted from the noisy biometric measurements. 
Assume further that the adversary is in the possession of another token $sp_{N+1}$, but not the template $u^{(N+1)}$, from the $(N+1)$'st
enrollment of the user. 

Let us denote by $A^{sp_i}$ the sets of matrices derived from the token $sp_i$. 
The adversary can either keep all corresponding sets of inequalities, or selectively choose the inequalities of  the  system  to  decrease  both  the  memory  usage  and the  running  time  to  refine  the  solution.
In the following, we denote by $\mathbf{\mathrm{\mathcal{A}}^{GRP}_{AC}}$  the attack consisting of using all the constraints, and by $\mathbf{\mathrm{\mathcal{A}}^{GRP}_{SC}}$ the attack where the constraints are selected.
The attack $\mathbf{\mathrm{\mathcal{A}}^{GRP}_{SC}}$ proceeds as follows:
\begin{enumerate}
    \item First, compute an approximated solution ${x^\prime}$ from the pair $(u^{(1)}, A^{sp_1})$, and
    initialize a set of constraints 
    \[
  S =\left\lbrace \langle \mathcal{W}^{sp_1}_{i,j} - \mathcal{W}^{sp_1}_{i,u^{(1)}_i}, x \rangle \leq 0 \;\middle|\;
  \begin{tabular}{@{}l@{}}
   $1 \leq i \leq m$\\ $1 \leq j \leq k$\\ $j \neq u^{sp_1}_i$
   \end{tabular}
  \right\rbrace.
\]
    \item For $b=2 \ldots N$, the following computations are performed:
    \begin{enumerate}
        \item ${u^\prime}= ({u^\prime}_1, \ldots, {u^\prime}_m)$ where ${u^\prime}_i=\argmax_{l\ \textrm{s.t.}\ 1 \leq l \leq k}  \langle \mathcal{W}^{sp_b}_{i,l},{x^\prime} \rangle\ \textrm{for}\ i=1 \ldots m$.
        \item $d={u^\prime} - u^{(b)}=(d_1, \ldots, d_m)$, a vector of differences.
        \item The set $S$ of constraints is updated as
        \[
        S \cup
        \left\lbrace \langle \mathcal{W}^{sp_b}_{i,j} - \mathcal{W}^{sp_b}_{i,u^{(b)}_i}, x \rangle \leq 0 \;\middle|\;
  \begin{tabular}{@{}l@{}}
   $d_i \neq 0$\\ $1 \leq i \leq m$\\ $1 \leq j \leq k$\\ $j \neq u^{(b)}_i$
   \end{tabular}
  \right\rbrace.
\]
        \item ${x^\prime}$ is updated subject to the constraints of $S$.
    \end{enumerate}
    \item Return ${x^\prime}$.
\end{enumerate}

    Recall that the dataset in \cite{Jin18Ranking} contains $5$ samples (genuine feature vectors) for each user. Therefore, in our experiments, we consider $2 \leq N \leq 4$. We use linear programming solver of the SciPy optimization library in Python. The \texttt{linprog} function is parameterized with the 'interior-point' solver method, with upper bounds (1) and lower bounds (-1) for the components of seeked solutions, and without objective function.
    The experiments yield the results of Table~\ref{IoM-GRP_TableMultipleLeaks_WO_Opts}  and Table~\ref{IoM-GRP_TableMultipleLeaks_SelectedConstraints}, showing an improvement of the comparison scores over the previous attacks (for $N=1$). Table~\ref{IoM-GRP_TableMultipleLeaks_WO_Opts} reports on the comparison scores obtained by an attacker $\mathrm{\mathcal{A}}^{GRP}_{AC}$; and Table~\ref{IoM-GRP_TableMultipleLeaks_SelectedConstraints} reports on the comparison scores obtained by an attacker $\mathrm{\mathcal{A}}^{GRP}_{SC}$, optimizing the number of constraints.

\begin{table}[hbtp!]
\centering
\caption{Comparison scores using $\mathrm{\mathcal{A}}^{GRP}_{AC}$.
}
\label{IoM-GRP_TableMultipleLeaks_WO_Opts}
\begin{tabularx}{\columnwidth}{l|X|X|X}
\toprule
$N$ & $2$   & $3$ & $4$  \\ 
\midrule
\makecell[l]{Constr. Number \\ ($N \cdot (k-1) \cdot m + 2 \cdot n$)} &  9,598  & 14,098 & 18,598  \\ 
GRP Comp. Score -- Min (\%) & 27.7  & 33   & 38   \\ 
GRP Comp. Score -- Avg (\%) & 50.6 & 53.4 & 56  \\ 
\bottomrule
\end{tabularx}
\end{table}

\begin{table}[hbtp!]
\centering
\caption{Comparison scores using $\mathrm{\mathcal{A}}^{GRP}_{SC}$.}
\label{IoM-GRP_TableMultipleLeaks_SelectedConstraints}
\begin{tabularx}{\columnwidth}{l|X|X|X}
\toprule
$N$ & $2$   & $3$ & $4$  \\ 
\midrule
Constr. Number -- Avg &  8,379  &  8,930  &  9,449  \\ 
GRP Comp. Score -- Min (\%) & 29.3   & 29.7   &   38.3 \\ 
GRP Comp. Score -- Avg (\%) & 48.4 & 50.6 &  52.8 \\ 
\bottomrule
\end{tabularx}
\end{table}

\subsection{Reversibility attacks on GRP-IoM}
\label{s: Rev GRP}

In authentication attacks in the previous section, adversarial strategies focus on finding nearby-template preimages $x^*$, that are not required to be close to the actual feature vector $x$.
In a reversibility attack,
an adversary finds a nearby-feature preimage $x^*$, and the quality of the attack is measured by the closeness of $x^*$ to $x$.

\paragraph*{\bf Exact reversibility}
The best case for an attacker is to have $x^*=x$. 
In \cite{Jin18Ranking},
it is argued that the best strategy for an attacker to find $x^*=x$ is to exhaustively search (guess) the components of $x$.
Given the feature vectors extracted from FVC2002-DB1, it is reported in~\cite{Jin18Ranking} that the minimum and
maximum values of the feature vector components are $-0.2504$ and
$0.2132$ respectively. Therefore, the search space for a feature component consists of $4636$ possibilities, including the positive and negative signed components. Moreover,
the fetaure vectors in GRP-IoM are of length $299$. Therefore, it is concluded in \cite{Jin18Ranking} that 
the attack requires to exhaust a search space of size
$4636^{299} \approx 2^{3641}$.
In the following, we propose a better attack strategy to recover $x$. The main idea is to guess the sign of the components of the feature vector, and shrink the search space accordingly. Given a token, template pair of a user, the adversary computes a nearby-template preimage
$x^*$, and guesses the sign of $x_i$ as the same as the sign of $x^*_i$. If all the signs were correctly guessed by the adversary, then the size of the search space would be reduced from $4636^{299}\approx 2^{3641}$ to $(4636/2)^{299}\approx 2^{3342}$. However, the adversary may guess the signs incorrectly. Based on our experiments, where we compare the sign of the components of the preimage vectors $x^*$ and the actual feature vectors $x$, we estimate that 
the probabilty of guessing the sign correctly per component is $242/299$. Therefore, we estimate the size of the search space for $x$ as
$(4636/2)^{299}\cdot(299/242)^{299}\approx 2^{3433}$.
Even though our attack is not practical, it reduces the previously estimated security level for GRP-IoM by $3641-3433=208$ bits from $3641$-bit to $3433$-bit.

\paragraph*{\bf Nearby reversibility}
Now, we analyze some attack strategies for finding a nearby-feature preimage of a template under the stolen token attack scenario. The adversary proceeds similarly as in the authentication attacks, except that now we also include some objective functions, and solve a linearly constrained quadratic optimization problem.
We consider three cases for which the objective functions are given as follows:
\begin{enumerate}
    \item $\min \|x\|^2_2$.
    \item $\min \|x - v_m\|^2_2$ where $v_m$ is the average feature vector in the database provided in~\cite{Jin18Ranking}. For our experiments, one sample per user is attacked, \emph{i.e.} one hundred linear programs are solved. 
    \item $\min \|x - v_r\|^2_2$ where $v_r$ is a feature vector derived from a fingerprint of the adversary. For our experiments, $v_r$ is picked at random among the samples of one user. These samples are then removed from the database. Among the remaining $99 \times 5$ samples, one sample per user is attacked, for a total of $99$ program solvings.
\end{enumerate}

In our experiments, we use Python and the CVXOPT package~\cite{cvxopt} which provides linearly constrained quadratic programming solvers. 
We measure the success rate of this attack strategy $\mathrm{Rate}^{rev}_{\mathcal{A}}$, and report its advantage over the false accept rate of the system.
We compute two reference  false accept rate values for the dataset provided in~\cite{Jin18Ranking}, one
with respect to the Euclidean distance,
and one with respect to the similarity measure as described in Remark~\ref{s: Different similarity measures}.
We compute $\mathrm{FMR}(\tau_{Euc})=\mathrm{FNMR}(\tau_{Euc})=0.03$ using the Euclidean distance, with the threshold $\tau_{Euc}=0.33$. 
We estimate $\mathrm{FMR}(\tau_{Sim})=\mathrm{FMR}(\tau_{Sim})=0.002$ using the similarity measure, with the threshold $\tau_{Sim}=0.13$.
Our attacks are evaluated in three cases: the cases~1, 2 and 3 when an objective function is used in the order as mentioned above, and the case~{\it none} when no optimization function is used. 
Results of our experiments, as summarized in Table~\ref{IoM-GRP_TableReversibility_Euclidean} and Table~\ref{IoM-GRP_TableReversibility_Special_Matching_Score}, show that in most cases
the solving of an optimization problem leads to a
$\mathrm{Rate}^{rev}_{\mathcal{A}}$ significantly greater that $\mathrm{FMR}(\tau_A)$. 
We then conclude that GRP-IoM is reversible with a single complete leak, both considering the Euclidean distance and the dedicated similarity score~of~\cite{Jin16Generating}.

\begin{table}[hbtp!]
\centering
\caption{Success rate of the reversibility attack against GRP-IoM under single stolen token and template attack; using the Euclidean distance and for different values of $\tau_A$.}
\label{IoM-GRP_TableReversibility_Euclidean}
\begin{tabularx}{\columnwidth}{l|X|X|X|X}
\toprule
Objective Function (case) & none &  1  & 2 & 3  \\ 
\midrule
$\mathrm{Rate}^{rev}_{\mathcal{A}}$, $\tau_{Euc}=0.33$ & 2 & 63  & 77  & 98   \\ 
$\mathrm{Rate}^{rev}_{\mathcal{A}}$, $\tau_{Euc}=0.2$ & 0 & 3  & 4  & 27.2   \\ 
$\mathrm{Rate}^{rev}_{\mathcal{A}}$, $\tau_{Euc}=0.15$ & 0 & 0  & 0  & 5   \\ 
$\mathrm{Rate}^{rev}_{\mathcal{A}}$, $\tau_{Euc}=0.1$ & 0 & 0  & 0  & 0   \\ 
\bottomrule
\end{tabularx}
\end{table}

\begin{table}[hbtp!]
\centering
\caption{Success rate of the reversibility attack against GRP-IoM under single stolen token and template attack; using the similarity measure with the threshold $\tau_{Sim}=0.13$.}
\label{IoM-GRP_TableReversibility_Special_Matching_Score}
\begin{tabularx}{\columnwidth}{l|X|X|X|X}
\toprule
Objective Function (case) & none &  1  & 2 & 3  \\ 
\midrule
$\mathrm{Rate}^{rev}_{\mathcal{A}}$, $\tau_{Sim}=0.13$ & 100 & 0  & 0  & 100    \\ 
\bottomrule
\end{tabularx}
\end{table}
Table~\ref{IoM-GRP_TableReversibility_Euclidean} also shows that an adversary's success rate drops when Euclidean distance threshold $\tau_{Euc}$ is lowered from $0.33$ in the system, as expected. 
In Table~\ref{IoM-GRP_TableReversibility_Special_Matching_Score}, we perform a similar analysis when the similarity measure is used with the threshold 
$\tau_{Sim}=0.13$. We observe that the best adversarial success rates are obtained when 
no objective function is used, or the objective
function in case 3 is used.
The effect of multiple stolen token and template pairs is evaluated for the previously mentioned four cases, the results of which are presented in the Table~\ref{IoM-GRP_TableReversibility_LeakMult_Case0_Euclidean} and~\ref{IoM-GRP_TableReversibility_LeakMult_Case2_Euclidean}. When no function is optimized, we see the success rate is increasing with the number of stolen pairs, up to 3 pairs, after which it decreases. This decrease may be due to the variability of the feature vector components at each re-enrollement of the user, \emph{i.e.} when a renewal of the token is required. Since each system of constraints that we add to the linear program corresponds to a re-enrollment, the amount of errors in the constants of the inequalities may exceed the benefit of having more inequalities.
Finally, when an objective function is added in the program solving, our experiments show there is no value gained with multiple leaks. We should note that
our experiments are rather limited due to the sample size. For better and more definitive conclusions, one would need to perform more experiments.

\begin{table}[hbtp!]
\centering
\caption{Success rate of the reversibility attack against GRP-IoM under $N$ stolen token and template attacks when no optimization is performed (case {\it none}), 
using the Euclidean distance and for different values of $\tau_A$.}
\label{IoM-GRP_TableReversibility_LeakMult_Case0_Euclidean}
\begin{tabularx}{\columnwidth}{l|X|X|X|X|X}
\toprule
$N$ & 1 & 2  & 3 & 4 & 5  \\ 
\midrule
$\mathrm{Rate}^{rev}_{\mathcal{A}}$, $\tau_{Euc}=0.33$ & 2 & 43  & 68  & 63 & 62 \\ 
$\mathrm{Rate}^{rev}_{\mathcal{A}}$, $\tau_{Euc}=0.2$ & 0 & 3  & 3  & 3 & 3  \\ 
$\mathrm{Rate}^{rev}_{\mathcal{A}}$, $\tau_{Euc}=0.15$ & 0 & 0  & 0  & 0 & 0  \\ 
\bottomrule
\end{tabularx}
\end{table}

\begin{table}[hbtp!]
\centering
\caption{Success rate of the reversibility attack against GRP-IoM under $N$ stolen token and template attacks when optimization is performed (case 2), using the Euclidean distance and for different values of $\tau_A$.}
\label{IoM-GRP_TableReversibility_LeakMult_Case2_Euclidean}
\begin{tabularx}{\columnwidth}{l|X|X|X|X|X}
\toprule
$N$ & 1 & 2  & 3 & 4 & 5  \\ 
\midrule
$\mathrm{Rate}^{rev}_{\mathcal{A}}$, $\tau_{Euc}=0.33$ & 77 & 71  & 69  & 69 & 69 \\ 
$\mathrm{Rate}^{rev}_{\mathcal{A}}$, $\tau_{Euc}=0.2$ & 4 & 3  & 3  & 3 & 3  \\ 
$\mathrm{Rate}^{rev}_{\mathcal{A}}$, $\tau_{Euc}=0.15$ & 0 & 0  & 0  & 0 & 0  \\ 
\bottomrule
\end{tabularx}
\end{table}

\section{Attacks on URP-IoM}
\label{s: Attacking URP}

In this section, we propose some concrete attack strategies against
URP-IoM, and evaluate the impact of our attacks through our implementation over one of the datasets as provided in \cite{Jin18Ranking}.
More specifically, we use the dataset of features 
extracted 
from the fingerprint images of FVC2002-DB1
as in ~\cite{Jin18Ranking}. This dataset contains a total of 500 samples:
5 samples per user for 100 users. 

As mentioned before,
for convenient comparison of our results, we use the parameter set $n=299$,
$(k,m,p)=(128, 600, 2)$ and $\tau=0.11$ as the main reference point in our security analysis, because these parameters are commonly referred in the security and performance analysis of URP-IoM in \cite{Jin18Ranking}; see Table~V in \cite{Jin18Ranking}.

\subsection{Authentication attacks on URP-IoM}
\label{s: Auth URP}

\paragraph*{\bf Finding nearby-template preimages} 


We let $x,u,sp$ as before, and assume that an adversary $\mathcal{A}$ knows $u$ and $sp$. In order to find a nearby-template preimage vector $x^*\in \mathbb{R}^{n}$, the adversary proceeds as follows.
Since $\mathcal{A}$ knows $sp$, $\mathcal{A}$ can recover the set of permutations $A=\{\sigma_{1,i}, \sigma_{2,i} \in S_{n,k}, 1 \leq i \leq m\}$ in URP-IoM.
The template $u=(u_1, u_2, \ldots, u_m) \in (\mathbb{Z}_k)^m$ is a vector comprised of the indices of maximum, \emph{i.e.} 
\[u_i=\argmax_{l\ \textrm{s.t.}\ 1 \leq l \leq k}  \{ x_{\sigma_{1,i}(l)} \cdot x_{\sigma_{2,i}(l)}\}\ \textrm{for}\ i=1 \ldots m,\]
where $k$ is the window size, from which the adversary can recover the set of inequalities 
\[
\left\lbrace x_{\sigma_{1,i}(j)} \cdot x_{\sigma_{2,i}(j)} \leq x_{\sigma_{1,i}(u_i)} \cdot x_{\sigma_{2,i}(u_i)} \;\middle|\;
  \begin{tabular}{@{}l@{}}
   $1 \leq i \leq m$\\ 
   $1 \leq j \leq k$\\
   $j \neq u_i$
   \end{tabular}
  \right\rbrace.
\]

Each of these inequalities can be transformed into linear constraints by taking the logarithm of the both sides. The corresponding set of $m\times (k-1)$ linear constraints can be given as follows:  
\[
\left\lbrace \begin{tabular}{@{}l@{}}
   $\log x_{\sigma_{1,i}(j)} + \log x_{\sigma_{2,i}(j)}$\\
   $- \left(\log x_{\sigma_{1,i}(u_i)} + \log x_{\sigma_{2,i}(u_i)}\right) \leq 0$
   \end{tabular} \;\middle|\;
  \begin{tabular}{@{}l@{}}
   $1 \leq i \leq m$\\ 
   $1 \leq j \leq k$\\
   $j \neq u_i$
   \end{tabular}
  \right\rbrace.
\]
The logarithm adds a new set of $n$ constraints, namely $x_i> 0$, $1\le i\le n$ where $n$ is the size of the feature vector.
$\mathcal{A}$ finds a solution $c^*$ of this system, and sets $x^*$ such that ${x_i}^* = exp({{c_i}^*})$, $i=1,...,n$, to be one of the (arbitrary) solutions of this system.
By the construction of 
$x^*$, we must have
$u^* = \Xi.\mathcal{T}(sp,x^*)=\Xi.\mathcal{T}(sp,x) = u$, and so 
$\Xi.\mathcal{V}(u,u^*,\tau_B) = True$, for all $\tau_B$.
In other words, $\mathcal{A}$ is expected to get (falsely) authenticated by the server with $100\%$, or equivalently, $\mathrm{Rate}^{Auth}_{\mathcal{A}}=100\%$.

The expected success rate of our attack
has been verified in our python implementation 
using the cvxopt library~\cite{cvxopt}
on a computer 
running on Ubuntu 17.10 with xfce environment,
with a Core i7 4790k 4Ghz processor, 8GB of RAM, and a SATA SSD of 512GB. The attack runs in the order of seconds for the parameters $n=299$,
$(k,m,p)=(128, 600, 2)$ and $\tau=0.11$.
We should note that, if an adversary captures more than one token and template pair, then additional constraints can further optimize the attack as previously discussed for GRP-IoM.


\paragraph*{\bf Finding long-lived nearby-template preimages}
We let $x,x^\prime,u,u^\prime,sp,sp^\prime$ be as before, and assume that an adversary $\mathcal{A}$ knows $u,sp, sp^\prime$.
In our proposed attack, $\mathcal{A}$ follows the previously described strategy to find a nearby-template preimage $x^*$ based on $u$ and $sp$, and presents this
$x^*$ as a candidate for nearby-template preimage of ${u^\prime}$.

We evaluate this attack by computing both the average and the minimum comparison score, over one hundred users, between ${{u^\prime}}^*=\Xi.\mathcal{T}({sp^\prime},x^*)$ and the re-enrolled genuine template ${u^\prime}=\Xi.\mathcal{T}({sp^\prime},{x^\prime})$. 
Our experiments yield $25.2\%$ as the  average rate of the number of indices with the same entry in ${u^\prime}$ and ${u^\prime}^*$;
and $3\%$ as the minimum rate of the number of indices with the same entry in ${u^\prime}$ and ${u^\prime}^*$. Therefore, given the comparison score thresholds of $\tau=0.11$ as set in~\cite{Jin18Ranking},
we expect that
the success rate of the adversary to be
$\mathrm{Rate}^{Auth-\ell\ell}_{\mathcal{A}}=100\%$, on average.
The above attack strategies show that URP-IoM is severely vulnerable against authentication attacks under the stolen token and template attack model, and also show that adversaries cannot be prevented by renewing templates or tokens. In other words, the cancellability feature of GRP-IoM is violated under the stolen token and template scenario.

\paragraph*{\bf Optimizing authentication attacks}

Similar to our GRP-IoM analysis, we now explore whether the attacks can be optimized when a user leaks several token and template pairs.
More specifically, assume that an adversary
captures $N$ token, template pairs 
$(sp_i,u^{(i)})$, for $1 \leq i \leq N$, derived from distinct feature vectors of the same user.
Assume further that the adversary is in the possession of another token $sp_{N+1}$, but not the template $u^{(N+1)}$, from the $(N+1)$'st
enrollment of the user. 

Table~\ref{IoM-URP_TableMultipleLeaks_WO_Opts} reports the values when the number of leaks increases, and shows that 2 stolen token and template pairs are sufficient to yield 
$\mathrm{Rate}^{Auth-\ell\ell}_{\mathcal{A}}=100\%$ when $\tau=0.11$.

\begin{table}[hbtp!]
\centering
\caption{Comparison scores using $\mathrm{\mathcal{A}}^{URP}_{AC}$.}
\label{IoM-URP_TableMultipleLeaks_WO_Opts}
\begin{tabularx}{\columnwidth}{l|X|X|X}
\toprule
$N$ & $2$   & $3$ & $4$  \\ 
\midrule
\makecell[l]{Constraint Number \\ ($N \cdot (k-1) \cdot m + 2 \cdot n$)} &  152,998  & 229,198  & 305,398  \\ 
URP Comp. Score -- Min (\%)  &  12.8  &  14.7 & 14.8 \\ 
URP Comp. Score -- Avg (\%)  & 28.2  & 29.6 & 31.3 \\ 
\bottomrule
\end{tabularx}
\end{table}

\section{Linkability Attacks on GRP-IoM and URP-IoM}
\label{s: linkability URP} 

Recall that an adversary $\mathcal{A}$ in a linkability attack can be modelled as an algorithm that 
takes $sp$, ${sp^\prime}$, $u$, and ${u^\prime}$ as input, and that outputs $0$ or $1$,
where the output $1$ indicates that 
the feature vectors $x$ and ${x^\prime}$ are extracted from the same user, and
the output $0$ indicates that 
the feature vectors $x$ and ${x^\prime}$ are extracted from two different users.

Authentication attacks on GRP-IoM and URP-IoM only return a feature vector that enables successful (false) authentication.
Reversibility attacks on GRP-IoM allows 
to construct a nearby-feature preimage vectors, that are somewhat 
close to the actual feature vector.
For example, in the exact reversibility attack on GRP-IoM,
we were able to guess the sign of a component of the actual 
feature vector with estimated probability of $242/299$.
In our linkability attack on GRP-IoM, we utilize such sign guessing, and partial reversibility results. 
However, we could not obtain nearby-feature preimage vectors in URP-IoM successfully, mainly because, by the use of geometric programming, all of the components
in a preimage must be non-negative, whereas an actual feature vector component can well be negative. As a result, the linkability attack techniques for GRP-IoM do not immediately apply to attack
URP-IoM. However, we show that it is still possible to successfully link URP-IoM templates.

\paragraph{An attack on GRP-IoM}
Given $sp$, ${sp}^\prime$, $u$, and ${u^\prime}$, the adversary computes nearby-feature preimage vectors $x^*$ and ${x'}^*$ as explained before.
For some  decision threshold value $t_{link}$, 
the adversary computes $\beta = \beta(x^*,{x'}^*)$, where 
$\beta$ is the number of indices
for which $x^*$ and ${x'}^*$ have exactly the
same sign.
Finally, the adversary outputs $1$, if $\beta\ge t_{link}$, indicating that
the feature vectors $x$ and ${x^\prime}$ are extracted from the same user.
Otherwise, if $\beta< t_{link}$, the adversary outputs $0$,
indicating that the feature vectors $x$ and ${x^\prime}$ are extracted from two different users.

In our experiments, we created 500 nearby-feature preimages, derived from the 500 templates along with their 500 seeds. Recall that the templates are the transformations (using distinct random seeds) of the feature vectors provided by the authors of IoM hashing~\cite{Jin18Ranking}.  Using our dataset of nearby-feature preimages (estimated feature vectors) produced by our attack, we estimate the success rate of our attack using the following script:
\begin{enumerate}
\item $c_1, c_2\leftarrow 0$
\item for $i$ between $1$ and $N$:
\begin{enumerate} 
\item pick at random two nearby-feature preimage vectors $x^*$ and ${x^\prime}^*$ from the same individual ($x^*\neq {x^\prime}^*$).
\item if $\beta(x^*,{x^\prime}^*)\geq t_{link} :$ $c_1 \leftarrow  c_1 + 1$. 
\item pick at random two nearby-feature preimage vectors $x^*$ and ${x^\prime}^*$ from two different individuals.
\item if $\beta(x^*,{x'}^*) < t_{link} :$ $c_2 \leftarrow  c_2 + 1$. 
\end{enumerate}
\item return $c_1/N$ and $c_2/N$.
\end{enumerate}
In our experiments, we set 
$t_{link}=170$ and $N=10000$,
and obtained
$c_1/N\approx 0.95$ and $c_2/N\approx 0.99$. Therefore, we estimate that $\mathrm{Rate}_{\mathcal{A}}^{Link}=0.97$,
as the average of the success rates 
over the genuine and imposter pairs.
The run time of the attack is dominated
by the run time of computing nearby-feature preimages, that takes only a few seconds as mentioned earlier.

\paragraph{An attack on URP-IoM}
Given $sp$, ${sp^\prime}$, $u$, and $u^\prime$, the adversary computes nearby-feature preimage vectors $x^*$ and ${x'}^*$ as explained before.
For some  decision threshold value $t_{link}$, 
the adversary computes 
the Pearson coefficient 
$\rho = \rho (x^*, {{x^\prime}}^*)\in [-1,1]$ of $x^*$ and ${{x^\prime}}^*$. 
The formula for the Pearson coefficient is given as follows:
\[
\rho(x,y)={\frac {\sum _{i=1}^{n}(x_{i}-{\bar {x}})(y_{i}-{\bar {y}})}{{\sqrt {\sum _{i=1}^{n}(x_{i}-{\bar {x}})^{2}}}{\sqrt {\sum _{i=1}^{n}(y_{i}-{\bar {y}})^{2}}}}},
\]
where $x=(x_1,...,x_n)$, $y=(y_1,...,y_n)$, $\bar{x} = \sum_{i=1}^{n}{x_i}/n$, and
$\bar{y} = \sum_{i=1}^{n}{y_i}/n$.

The adversary outputs $1$, if $|\rho|\ge t_{link}$, indicating that
the feature vectors $x$ and ${x^\prime}$ are extracted from the same user.
Otherwise, if $|\rho|< t_{link}$, the adversary outputs $0$,
indicating that the feature vectors $x$ and ${x^\prime}$ are extracted from two different users.
Following the linkability attack on GRP-IoM, we estimate the success rate of our attack using the previous script, but replacing $\beta (x^*, {x^\prime}^*)$ by $\rho (x^*, {x^\prime}^*)$.
In our experiments, we set 
$t_{link}=0.18$ and $N=10000$,
and obtained
$c_1/N\approx 0.83$ and $c_2/N\approx 0.83$. Therefore, we estimate that $\mathrm{Rate}_{\mathcal{A}}^{Link}=0.83$,
as the average of the success rates 
over the genuine and imposter pairs.
The run time of the attack is dominated
by the run time of computing nearby-feature preimages, that takes only a few seconds as mentioned earlier.








\section{Conclusion}
\label{s: Conclusion}
We formalized the authentication, irreversibility and unlikability notions under the stolen token scenario, and proposed several attacks against GRP-IoM and URP-IoM. We argued that both schemes are severely vulnerable against authentication and linkability attacks. Based on our experimental results, we estimated $100\%$ success rate for our authentication attacks against GRP-IoM and URP-IoM, 
$97\%$ success rate for our linkability attacks against GRP-IoM,
and $83\%$ success rate for our linkability attacks against URP-IoM.
We also proposed better reversibility attacks against GRP-IoM, but they are not practical yet. 


We believe that our attacks can further be improved. One interesting research direction would be to see the impact of different choices of objective functions in modelling the optimization problems in the authentication and reversibility attacks. Similarly, it would be interesting to exploit different correlation metrics in the linkability attacks.

Finally, we assume that adversaries are not adaptive and they are not allowed to ask queries for data of their choices in our attack models. This is rather a weak adversarial model. Therefore, we expect that our attacks can further be improved by allowing stronger adversaries. 

 \section{Acknowledgements}
This work was initiated when Koray Karabina was visiting the GREYC lab, ENSICAEN, CAEN, France. This material is based upon work supported by the National Science Foundation under Grant No. (NSF-CNS-1718109).
A part of this work was completed when Kevin Atighehchi and Loubna Ghammam were working at the GREYC lab.

\bibliographystyle{IEEEtran}
\bibliography{biblio}

\end{document}